\documentclass[10pt,conference]{IEEEtran}
\IEEEoverridecommandlockouts
\usepackage[utf8]{inputenc}
\usepackage{ae}
\usepackage{booktabs}
\usepackage{amsmath}
\usepackage[cmintegrals]{newtxmath}
\usepackage[noadjust]{cite}
\usepackage{multirow}
\usepackage{caption} 
\usepackage{url}
\usepackage{xspace}
\usepackage{xcolor}
\definecolor{alexpink}{HTML}{FFA0A0}
\usepackage[color=oxygenorange]{todonotes}

\usepackage[breaklinks]{hyperref}
\usepackage{breakurl}
\title{\sysname: Scalable Catalog Delivery in Privacy-Preserving Advertising}

\author{
{\rm Muhammad Haris Mughees}*\thanks{*Work was done whilst author was an
intern at Brave Software.} \\ 
    University of Illinois\\ Urbana-Champaign
\and
{\rm Gonçalo Pestana} \\ 
    Brave Software
\and
{\rm Alex Davidson} \\ 
    Brave Software
\and
{\rm Benjamin Livshits} \\ 
    Brave Software \\
    Imperial College London
}

\newcommand{\ith}[1]{\ensuremath{#1^{\text{th}}}\xspace}
\newcommand{\sysname}{\textsf{PrivateFetch}\xspace}
\newcommand{\pir}{\textsc{PIR}\xspace}
\newcommand{\pirscheme}{\ensuremath{\mathsf{PIR}}\xspace}
\newcommand{\init}{\ensuremath{\mathsf{Init}}\xspace}
\newcommand{\query}{\ensuremath{\mathsf{Query}}\xspace}

\newcommand{\reply}{\ensuremath{\mathsf{Reply}}\xspace}
\newcommand{\extract}{\ensuremath{\mathsf{Extract}}\xspace}

\newcommand{\simhash}{\ensuremath{\mathsf{SimHash}}\xspace}

\newcommand{\state}{\ensuremath{\mathsf{st}}\xspace}
\newcommand{\db}{\ensuremath{\mathsf{DB}}\xspace}

\newcommand{\floc}{\textsc{FLoC}\xspace}

\newcommand{\point}[1]{\par\smallskip\noindent{\bf{#1.}}~}

\newcommand{\AdCount}{\ensuremath{30}\xspace}
\newcommand{\DBSize}{\ensuremath{>1,000,000}\xspace}
\newcommand{\ClientCount}{\ensuremath{10,000,000}\xspace}
\newcommand{\Runtime}{\ensuremath{40}\xspace}
\newcommand{\Comms}{\ensuremath{192}\xspace}
\newcommand{\DollarsSaved}{\ensuremath{>\$3}million\xspace}

\newcommand{\adcatalog}{\ensuremath{\mathsf{AdCatalog}}\xspace}
\newcommand{\indextree}{\ensuremath{\mathsf{IndexTree}}\xspace}

\newcommand{\pparams}{\ensuremath{\mathsf{pp}}\xspace}

\begin{document}
\sloppy

\maketitle

\begin{abstract}
    A privacy-oriented recalibration of the Internet (e.g., by removing
    traditional tracking vectors like third-party cookies) is likely to
    drive a commodification of Internet assets, content, and
    infrastructure. As a result, users are expected to have to cover the
    revenue shortfalls themselves.
    
    In order to preserve the possibility of an Internet that is free at
    the point of use, attention is turning to new solutions that would
    allow targeted advertisement delivery based on behavioral
    information such as user preferences, without compromising user
    privacy. Recently, explorations in devising such systems either take
    approaches that rely on semantic guarantees like
    \(k\)-anonymity~---~which can be easily subverted when combining
    with alternative information, and do not take into account the
    possibility that even knowledge of such clusters is privacy-invasive
    in themselves. Other approaches provide full privacy by moving all
    data and processing logic to clients~---~but which is prohibitively
    expensive for both clients and servers.
    
    In this work, we devise a new framework called \sysname for building
    practical ad-delivery pipelines that rely on cryptographic hardness
    and best-case privacy, rather than syntactic privacy guarantees or
    reliance on real-world anonymization tools. \sysname utilizes local
    computation of preferences followed by high-performance
    single-server private information retrieval (PIR) to ensure that
    clients can pre-fetch ad content from servers, without revealing any
    of their inherent characteristics to the content provider. When
    considering an database of~\DBSize ads, we show that we can
    deliver~\AdCount ads to a client in~\Runtime seconds, with total
    communication costs of~\Comms{}KB. We also demonstrate the
    feasibility of \sysname by showing that the monetary cost of running
    it is less than $1\%$ of average ad revenue. As such, our system is
    capable of pre-fetching ads for clients based on behavioral and
    contextual user information, before displaying them during a typical
    browsing session.

    In addition, while we test \sysname as a private ad-delivery, the
    generality of our approach means that it could also be used for
    asynchronous and private fetching of other content types with
    minimal changes to the protocol flow.
\end{abstract}

\section{Introduction}
\label{sec:intro}
Most of the services and applications deployed on the web require
dynamic interactions between clients and servers
maintained by first- and third-party providers in order for clients to
display content to the user. Such interactions include, amongst others,
advertisement delivery for online advertising
(OA)~\cite{FLOC,BraveAdCatalog} applications, checking of certificate
revocation in TLS~\cite{SP:LCLMMW17}, checking of compromised login
credentials~\cite{CCS:LPASCR19}, and any other content that is requested
directly by the user. In almost all situations, the client is required
to provide information that reveals certain aspects of the user profile
to the content provider.

To enable behavioral-based targeting in OA, users are expected to provide subtle characteristics related to
their own preferences and browsing history. Revealing such
characteristics significantly compromises their privacy. However, devising an
Internet devoid of such advertising would inevitably become an
expensive place to inhabit for users.
Spending in OA markets was valued at~378 billion USD in
2019~\cite{OASpend}, whilst Google and Facebook were estimated to depend
on advertising for 83\% and 99\% of their revenue,
respectively~\cite{IAB,WiredEdelman,ForbesGoogle}. Clearly, website
operators would lose a huge chunk of their income.

Thus, with the OA industry likely to stay, the focus has turned to
trying to reimagining it in a way that it is possible to deliver
ad-based content to client devices without compromising user privacy.
While some approaches have considered only targeting ads based on
contextual information (such as the current webpage that is being
viewed), such systems do not give high utility to advertisement
providers~\cite{EPRINT:SerHogDev21}. Therefore, finding new private
solutions that allow behavioral targeting based on inherent user
preferences is likely to lead to much more relevant ads, and
subsequently higher user engagement.

Unfortunately, existing approaches to maintain private behavioral
targeting either end up delivering entire databases to clients to
perform all computation locally~---~which usually results in huge
bandwidth and performance costs~\cite{BraveAdCatalog}~---~or relying on
semantic definitions, such as \(k\)-anonymity, where user privacy
is both much harder to quantify, and vulnerable to compromise if the
clusters themselves leak too much information~\cite{FLOC}. To bridge the
gap between ensuring user privacy, whilst maintaining the utility and
low cost of Internet usage, we design and implement \sysname.

\begin{figure}[t]
    \centering
    \includegraphics[width=\columnwidth]{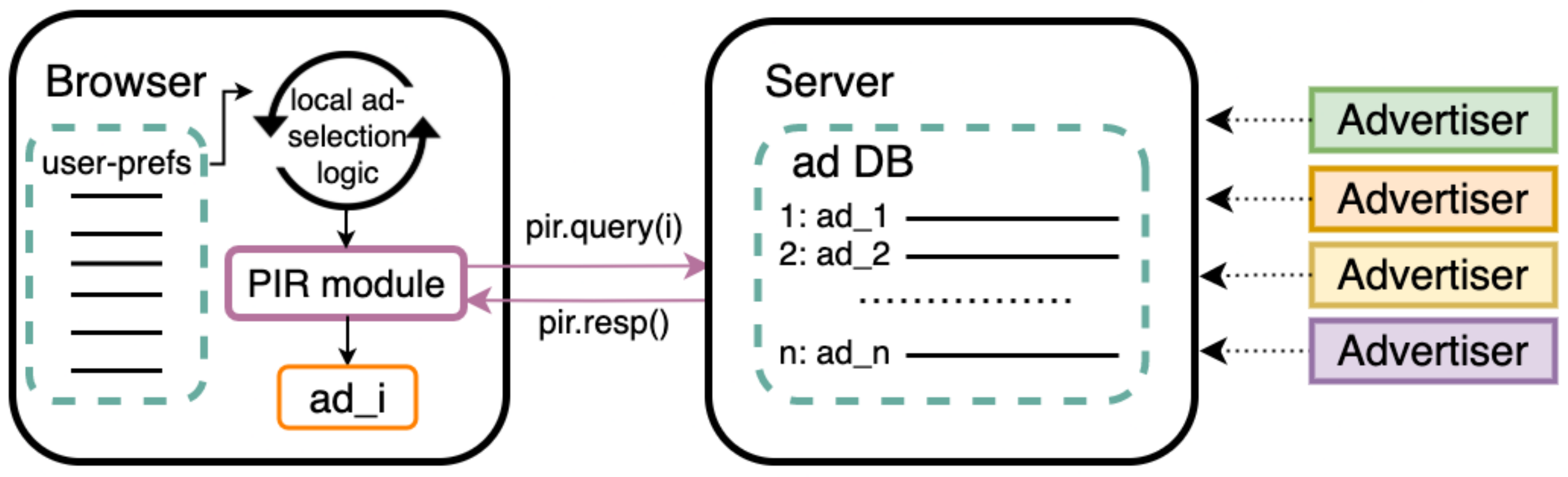}
    \caption{Overview of \sysname design for ad-delivery.}
    \label{fig:pir-ads-overview}
\end{figure}

\point{Overview of \sysname}
\sysname is a practically efficient, \emph{privacy-preserving}
advertisement targeting and delivery system that is intended for
deployment in client browsers and a wide range of web applications.
Advertisement targeting is done locally on the client device using
commonly-used tools for locality sensitive hashing, based on contextual
and behavioral data obtained from the client. Once a client's targeted
ads are determined, the client then retrieves them from the untrusted
proxy that handles the advertisement database via a private information
retrieval (PIR) protocol~\cite{FOCS:KusOst97}. See
Figure~\ref{fig:pir-ads-overview} for a general overview of the system.

The technical challenge of our system is ensuring that a client's ad
preference is not leaked to the ad provider, while at the same time
ensuring that the system has a high rate of conversions and keeping the
overall monetary cost of running the system minimal. The usage of a PIR
protocol ensures that \emph{nothing} about the client query is revealed
to an honest-but-curious advertisement proxy. Additionally, we highlight
the applicability of our system by devising a framework for building a
useful, cost-effective, and high-performance advertisement delivery
network for all participants. We can choose an underlying PIR scheme
such that \sysname can deliver~\(30\) advertisements to a client
in~\(10\) seconds from a database of~\(>250,000\) advertisements. For
instance, this allows us to build applications that allows clients to
asynchronously query advertisements based on behavioral and contextual
browsing, and then displaying ads to the user without having to wait for
webpages for a prohibitively long time. Moreover, this is all achieved
without revealing any of the private characteristics or preferences of
the client that are used in ad targeting to the ad-delivery server
(beyond what is revealed in higher layers, such as in a standard HTTPS
request).

Overall, \sysname is orders of magnitude cheaper than delivering the
database to each client locally and performance is similar to existing
solutions that still require some information to be leaked to the
proxy~\cite{EPRINT:SerHogDev21,NSDI:GuhCheFra11,NDSS:TNBNB10}. We
achieve this by making modifications to state-of-the-art single-server
PIR protocols that make using such schemes in our setting viable. Such
protocols are notoriously heavy on computation, and so we devise a
bucketization mechanism that allows clients to retrieve multiple
advertisements in a single PIR query based around locality-sensitive
hashing techniques~\cite{FLOC}. Furthermore, unlike previous systems,
our architecture does not require hardware support~\cite{SP:BKMP12},
centralization of client preference processing~\cite{JC:BeiIshMal04},
reliance on external anonymization networks~\cite{EPRINT:SerHogDev21},
or higher client-side storage and computation~\cite{NDSS:TNBNB10}.

Finally, while \sysname is primarily targeted towards the
OA use-case, we believe that the fetching model may be of use to other
applications. In essence, \sysname provides the capability for private
and asynchronous fetching of indexed content in the Internet setting.

\subsection{Our Contributions}
The formal contributions in this work follow.
\begin{itemize}
    \item \emph{A generic and modular framework for fully private
    advertisement delivery}: We develop a generic framework, \sysname,
    for online advertisement delivery to clients, whilst maintaining
    \emph{absolute privacy} in the honest-but-curious model: no locally
    computed preferences are ever learnt by the content provider.
    Our framework is generic in that it can be instantiated using any
    PIR scheme. The generality of our approach may have independent
    value in other content-delivery scenarios.
    
    \item \emph{Optimizations in Private Information Retrieval}: For our
    application, we choose to implement \sysname using the
    OnionPIR~\cite{EPRINT:MugCheRen21} stateless single-server PIR
    scheme, that is based on SEAL Fully Homomorphic
    Encryption~\cite{SEAL}. We improve the OnionPIR scheme by having the
    server bucketize their ad database relative to a locality sensitive
    hash function such as \simhash~\cite{FLOC}. This enables multiple
    ads for the same category to be retrieved using a single PIR query.
    In order to make this work, we must utilize a property of the
    OnionPIR scheme to embed multiple PIR responses into a single FHE
    ciphertext.
    
    \item \emph{Practical implementation of end-to-end protocol}: To
    demonstrate performance and applicability, we implement \sysname and
    show that we improve on the naive end-to-end system of delivering
    entire ad catalogs by several orders of magnitude, as well as being
    comparable to solutions that result in categorically higher leakage
    profiles. Overall, \sysname can deliver~\AdCount ads within~\Runtime
    seconds to a client when considering an ad database with~\DBSize
    entries. These improvements translate into very low financial
    running costs: \DollarsSaved saved compared with the naïve approach,
    rendering it practical for everyday use.
\end{itemize}

\subsection{Limitations}

\point{Reporting conversions and other metrics}
In this work, we only cover the targeting and delivery portion of a full
advertisement network. Recognizing that advertisement networks are
usually driven by determining \mbox{click-through} rates for each ad, we note
that it is necessary for clients to report which ads it has interacted
with back to the network, and that this should be done whilst
maintaining their privacy. In Section~\ref{sec:discussion}, we discuss
how it is possible for our delivery pipeline to be integrated as a
generic module into many existing privacy-preserving advertisement
reporting
mechanisms~\cite{ARXIV:PQPL21,EPRINT:SerHogDev21,AppleConversions}.

\point{Real-time advertisement auctions} Our construction is unable to
support real-time advertisement auctions. Unfortunately, practical
systems supporting this functionality do so at the cost of revealing
which ads are queried for~---~even if the link to which client has
retrieved them is not
maintained~\cite{EPRINT:SerHogDev21}.\footnote{Revealing such
information may still be enough to produce linkability between client
profiles and queries, depending on what other public information is
known outside of the system.} While there are advantages in enabling
such functionality, we focus in this work on providing \emph{absolute
privacy} for client queries (i.e., not even revealing to the server
which ads are queried). See Section~\ref{sec:discussion} for a more
detailed discussion.

\section{Background}
\label{sec:background}
\subsection{Online Advertising}
In online advertising (OA), most systems first profile and track users
based on their actions across different websites and then display
relevant ads to the targeted user based on their \emph{behavior} and
other \emph{contextual} information. Specifically:

\begin{itemize}
    \item \emph{behavioral} ad targeting requires matching ads to users
    based on their (usually private) preferences;
    \item \emph{contextual} ad targeting is performed relative only to
    the actions performed by a user and relative to the content that
    they request.
\end{itemize}
The advantage of contextual ad targeting compared with behavioral is
that the targeting mechanism does not rely on any private user data.
However, ads are demonstrably less relevant in the contextual setting,
since behavioral-based approaches can be used to solicit ads that are
likely to match a given user's profile. In principle, for any targeting
mechanism to guarantee \emph{high utility}, it must use both contextual
and behavioral data in order to display the most relevant ads. Such
utility is usually measured in terms of an \emph{ad conversion rate} or
\emph{click-through rate}. This rate measures the ratio of users that
interact with each advertisement.

Clearly, performing behavioral targeting comes at the expense of a
user's privacy. A user's behavioral traits, when revealed either alone
or combined with other garnered or public information, reveal a
non-quantifiable amount of information about a user's personality,
traits, and habits. It goes without saying that such information should
be kept private from third-parties at all times. A user's right to
ensuring their data privacy is one of the key cornerstones of the
European General Data Protection Regulation~(GDPR).

As a result, the OA space has seen a number of recent innovations that
attempt to maintain the lucrative business of Internet advertising,
whilst providing much better privacy guarantees for clients. Overall,
ad-delivery systems should satisfy the following requirements:
\begin{itemize}
    \item \emph{Full privacy}: All client preferences that are used for
    retrieving the appropriate ad content for the user should be kept
    private from the content provider.
    \item \emph{High utility content}: Clients should receive
    advertisements that reflect their own private characteristics and
    interests.
    \item \emph{Minimal delivery latency}: Clients should receive
    \mbox{ad-content} in a timely manner so that the content is
    \emph{fresh}, and \emph{relevant} (with regards to their changing
    preferences).
    \item \emph{Low bandwidth usage}: Bandwidth usage should be kept at
    a minimum to keep server costs low, since such systems are likely to
    be used by very large numbers of clients.
    \item \emph{High rate of database updates}: Due to the nature of ad
    auctions, the system should tolerate a rapidly changing server-side
    database.
    \item \emph{Configuration flexibility}: The building blocks of the
    framework should ideally allow configuring components for varying
    efficiency/functionality trade-offs. This flexibility should be
    supported by the cryptographic primitives that are used.
\end{itemize}

\subsection{Threat Model}
\label{sec:threat}

In \sysname, we assume that proxy server and advertisers are
\emph{honest-but-curious}. Specifically, we assume that the proxy server
does not alter the catalog of the ad on its own, alter or reject the
client query, or hold the delivery of ad after query. Similarly, we assume
that the advertisers upload valid ads that they want to display to the
clients.

Unlike other \mbox{multi-party} computation systems, \sysname does not require the
proxy server and the advertisers to be \mbox{non-colluding}. The security of
\sysname holds even if the proxy server is run by the advertiser. 

Though it is possible to generically compile \sysname into a maliciously
secure system~\cite{STOC:GolMicWig87}, generally such systems suffer from
high overhead and multiple rounds of communications. Thus, we briefly
explore some practical mechanisms for ensuring robust deployments
against a proxy that attempts to act maliciously in
Section~\ref{sec:discussion}.

\subsection{Comparable approaches}
The desire for private advertisement targeting and delivery has been
discussed in academic literature across various points in the last
decade~\cite{NSDI:GuhCheFra11,NDSS:TNBNB10,ARXIV:PQPL21,EPRINT:SerHogDev21}.
However, in terms of practical deployments, there are few examples of
systems that have been trialled and used~\cite{FLOC,BraveAdCatalog}.
Below, we discuss two particular ad-delivery pipelines (and targeting
philosophies) that have been developed with the goal of providing
advertisements to client devices, whilst keeping their contextual data
private.

\point{Google \floc}
\begin{figure}[t]
    \centering
    \includegraphics[width=\columnwidth]{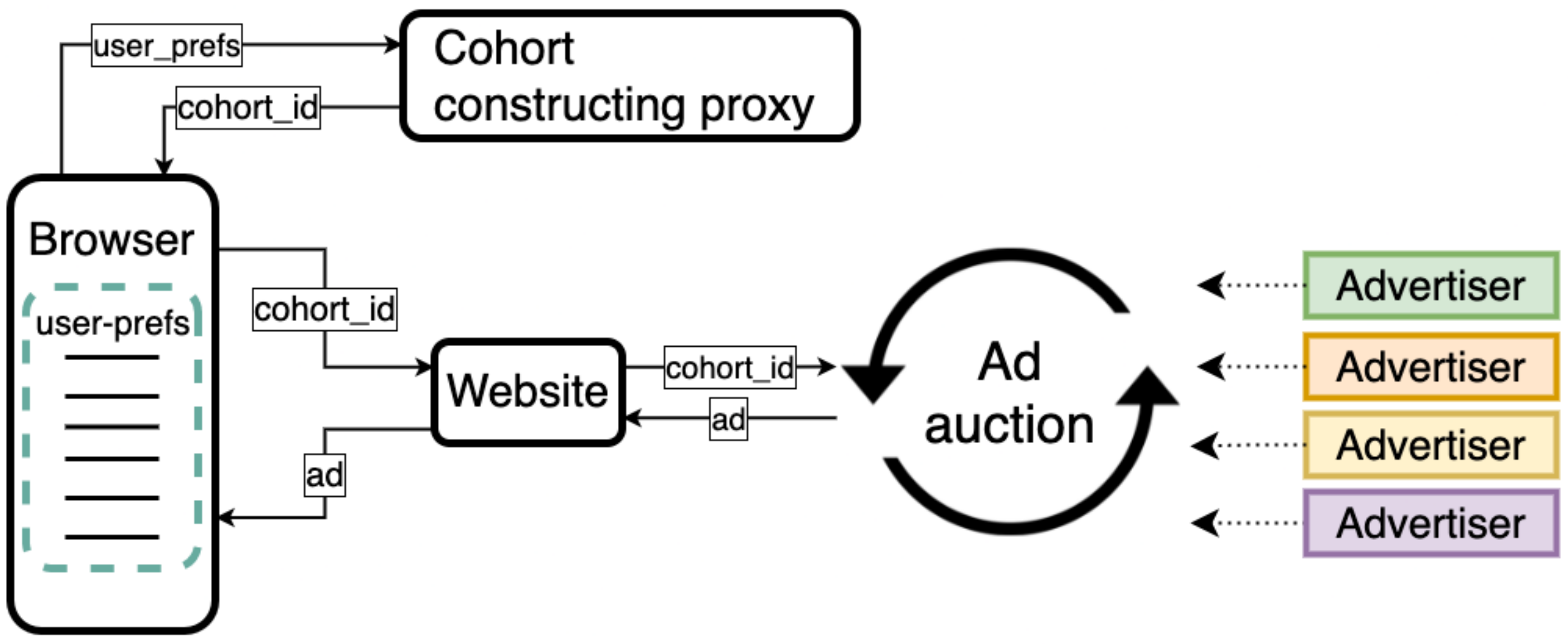}
    \caption{General idea behind Google \floc approach~\cite{FLOC}. Note
    that cohorts are assembled based on various indicators such as user
    preferences and device characteristics~\cite{FLOCWhitePaper}.}
    \label{fig:floc-overview}
\end{figure}
In 2020, Google proposed a new system called Federated Learning of
Cohorts (\floc)~\cite{FLOC}.\footnote{Following extensive analysis of
privacy pitfalls related to \floc~\cite{MozillaFLOC}, the expected
roll-out has since been delayed until late 2023~\cite{FLOCdelay}.} At a
high level, the system constructs \emph{cohorts} of users that share a
combined set of interests and preferences
(Figure~\ref{fig:floc-overview}). In order to develop such cohorts, the
browser keeps track of the browsing history of the user, and also
inherent characteristics (such as the user device) and assigns the user
to one of the global cohorts. As the user browsers the web, the browser
shares the user cohort with websites and advertisers. The advertisers
can then use this information to show relevant ads to the user. To
maintain privacy each cohort must contain a pre-determined and large
number of users. The idea is that, within the cohort, the user remains
anonymous, and so they receive privacy guarantees essentially amounting
to \(k\)-anonymity. A central administrator or proxy counts the number
of users in each cohort and, if required, merges smaller relevant
cohorts. 

Even though the approach utilized in \floc would appear to be robust
against existing privacy issues in OA via third-party tracking, it
induces new issues of its own~\cite{MozillaFLOC}. Firstly, sharing
cohort IDs with the advertisers allows the advertisers to enough
information about users to derive sensitive information about the user's
character anyway. Secondly, it has been shown that even learning benign
information about a user, such as movie reviews, can be enough to learn
their personal and sensitive information such as political ideologies.
Thirdly, when combining this information with other publicly available
information, the privacy loss is likely to be dramatically worse.
Finally, the user's privacy is entirely dependent on the cohort that
they fit into and this management is handled by a centralized proxy. As
a result, \floc's pitfalls are liable to give advertisers and trackers
information significant identifying and sensitive information about a
user's profile~\cite{MozillaFLOC}.

\point{Local preference computation}
\begin{figure}[t]
    \centering
    \includegraphics[width=\columnwidth]{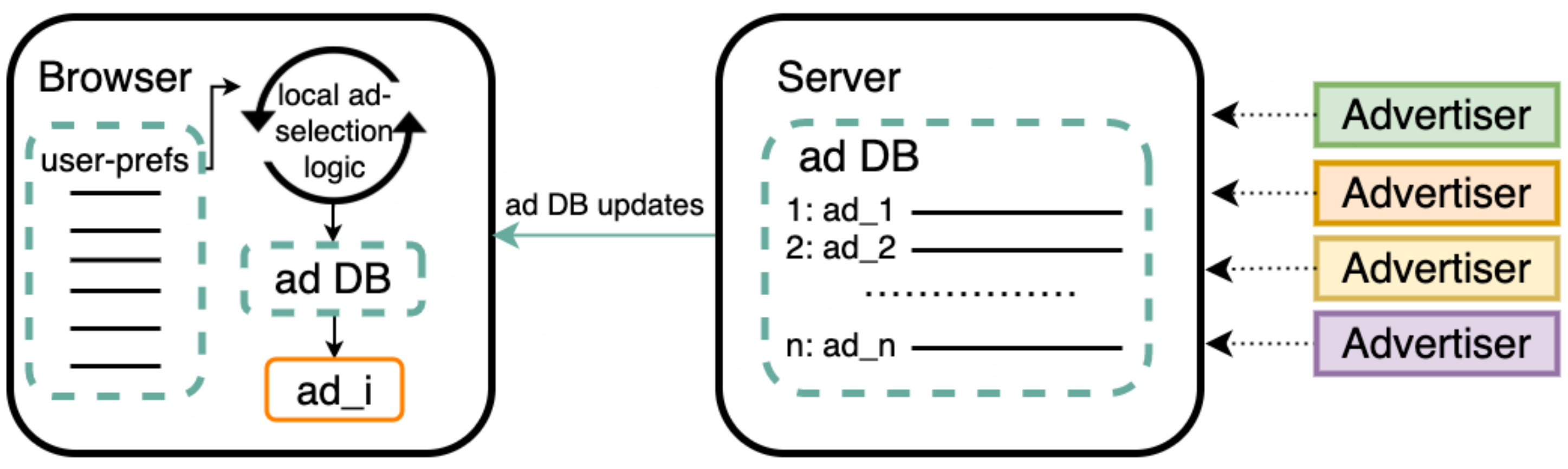}
    \caption{Overview of a naïve system that simply provides the entire
        advertisement database to each client. Such systems are already
        used widely in production by millions of Internet
        clients~\cite{BraveAdCatalog}.}
    \label{fig:local-ads-overview}
\end{figure}
Building web services that do not reveal user-specific traits to
websites and third-party requires a combination of local computation,
client storage, and a private content delivery mechanism. A simple
approach to achieve privacy preserving content delivery is to download
the full database from the service provider and selecting the desired
items locally (Figure~\ref{fig:local-ads-overview}). Such systems
preserve privacy since the user does not leak to third parties any
information as to which content it is requesting from the database.
However, these systems require very large bandwidth overheads as the
client must fetch a database that is potentially gigabytes in
size.\footnote{For example, assuming that each client downloads the ad
an ad database of approximately \(1\)GB (\(1\)KB ads \(\times\)
\(1,000,000\) entries).}

Unfortunately, without any better privacy-preserving alternatives many
large scale systems have adopted this approach and pay huge bandwidth
price.  
Most notably, projects such as the Brave
browser\footnote{\url{https://brave.com/brave-ads/}}, have been forced
to turn to deliver entire ad catalogs to users in order to maintain
privacy preserving ad-delivery functionality~\cite{BraveAdCatalog}. Such
mechanisms are almost impossible to maintain as the advertisement
database grows, as well as the number of clients and frequency of
profile updates increase.

\subsection{Private Information Retrieval}
\label{sec:pir-background}
We use Private Information Retrieval (PIR) as a generic functionality
for retrieving advertisements from an untrusted server. To guarantee
that the system can be used in the Internet setting, we must choose the
PIR protocol carefully to ensure that runtimes, bandwidth, and
associated financial costs are kept to a minimum. While numerous
advancements have been made in recent PIR literature, many schemes still
have unacceptable overheads for our application. We discuss the reasons
behind which scheme we choose here, and provide a more detailed
background on the state-of-the-art PIR literature in
Section~\ref{sec:pir-related}.

\point{Stateless single-server PIR}
In prior single-server \emph{basic} PIR schemes to generate a response,
the server has to perform a linear amount of computation over the
database~\cite{SP:ACLS18,EPRINT:MugCheRen21}. Additionally, these
schemes are based on computational cryptographic assumptions. Therefore
the server has to perform at least one cryptographic operation per
database element. These schemes are usually the easiest to deploy,
requiring no management of state or database updates, and no trust
assumptions to be made by the client. Furthermore, while such schemes
usually have relatively high computational overhead (compared to
two-server or stateful PIR schemes), they are still relatively
practical. For example, to retrieve an element from a database with one
million entries using the OnionPIR scheme~\cite{EPRINT:MugCheRen21}, it
takes~$\sim{40}$ seconds and less than \(200KB\) of communication.

\point{Stateful single-server PIR}
\emph{Stateful} single-server PIR scheme reduces computational
overhead~\cite{CCS:PatPerYeo18}. Specifically, the client first
interacts with the server in the offline phase to retrieve a
\emph{hint}. Then in the online phase, the client can use the hint to
perform \emph{cheaper} PIR queries. As a result, the server has to
perform only a sub-linear number of online expensive cryptographic
operations\footnote{The server still performs a linear number of PRF
evaluations}. The major drawback of this scheme is that to retrieve a
hint, the client has to download the whole database locally. Each hint
allows making a bounded number of cheap online queries, therefore the
client also has to perform this step repeatedly.

\point{Two-server PIR}
In two-server (or generally information-theoretic) PIR schemes, the
database is replicated on two \emph{non-colluding} servers. The server
computation does not involve performing any cryptographic operation.
Therefore, these schemes are relatively cheaper than the single-server
variants. \emph{Stateful} two-server PIR schemes have further reduced
the amortized server computation~\cite{EC:CorKog20}. Therefore, in terms
of computation, these schemes are ideal for applications where low
latency is required. However, the assumption of two or more
non-colluding servers makes these schemes unsuitable for many practical
applications.

\point{Final choice of scheme}
Comparing each of the available schemes, we find that two-server schemes
introduce unacceptable overheads in terms of both implementation and
running complexity, since such schemes require non-colluding servers
that both process PIR queries. Therefore, we focus only on single-server
schemes.

While stateful single-server schemes enjoy amortized performance
advantages over stateless schemes, we find the bandwidth costs
prohibitive for setting up a production OA system. Such costs arise due
to the need to send large portions of the ad database clients in the
initial step. Moreover, such schemes suffer dramatically when
considering the potential for database updates that render previous
client state redundant. In such cases, such an approach would require
careful implementation of the necessary fresh downloads of state.

With these concerns in mind, we turn to the state-of-the-art in
single-server PIR, known as OnionPIR~\cite{EPRINT:MugCheRen21}. We find
that this scheme is performant enough for our application, when compared
with existing approaches~\cite{FLOC,BraveAdCatalog}. OnionPIR requires a
server to perform 10 seconds of computation per ad query (for an ad
database of \(250,000\) elements), where each client is likely to make
10 queries per 3 hour period~\cite{BraveAdCatalog}. Such querying can be
performed asynchronously to client browsing, based on an up-to-date
locally computed approximation of the client's interests and
preferences.

In summary, we implement our private ad-delivery system on top of the
OnionPIR scheme~\cite{EPRINT:MugCheRen21}. However, we stress that our
approach is generic and that the explicit choice of PIR scheme can be
made independently, in order to optimize the overall system for the
desired application. We provide a further examination of using
alternative PIR schemes in Section~\ref{sec:discussion}.

\subsection{Overall PIR Framework}
\label{subsec:pir_framework}

As mentioned above, we only consider stateless single-server PIR. Such a
scheme involves a client with a secret index \(i\), and a server holding
a database \(\db\), that the client wants to learn the \(\ith{i}\)
record from. The query process of the PIR protocol consists of a single
message from the client to the server, and a single response from the
server to the client. The client then uses this response to recover the
\ith{i} record. Formally, PIR is defined using the following algorithms.

\begin{itemize}
    \item $(\pparams) \gets \pirscheme.\init(c, \db)$: A protocol
    executed by the client and the server. The client takes as input
    parameter \(c\), describing number of records that can be stored by
    the client. The server takes database $\db$ as input. The client's
    output is the initial state $\state$ and server gets no output. 

    \item $(q) \gets \pirscheme.\query(j,\state)$: An algorithm that is
    executed by the client. It takes as input index $j \in [n]$ and
    current state $\state$ and outputs an encrypted query $q$ to be sent
    to the server.
    
    \item $(r) \gets \pirscheme.\reply(q, \db)$: An algorithm that is
    executed by the server. It takes as input an encrypted query \(q\)
    and the database \(\db\), and outputs an encrypted reply $r$.
    
    \item $(\mathsf{e}_j) \gets \pirscheme.\extract(r, \state)$ An
    algorithm that is executed by the client to extract a record from
    server's reply.
\end{itemize}

For any PIR protocol, we have the followings requirements.
\begin{itemize}
    \item \textsc{Correctness}: Informally, the correctness of PIR
    requires that the client that queries~\(j\) learns the desired entry
    $\mathsf{e}_j$ with high probability.
    \item \textsc{Security}: The security of PIR requires that the
    server learns no information about the clients queried index~$j$.
\end{itemize}

\subsection{Locality-Sensitive Hashing}
In \sysname, we use locality-sensitive hashing (LSH) to enable batching
multiple advertisements of the same category together. This increases
performance of our system by allowing clients to retrieve multiple ads
that are likely to be interesting to them in a single PIR query.

To implement this mechanism we use \simhash, a LSH function that hashes
similar items into the same \emph{buckets} with high
probability~\cite{STOC:Charikar02}. \simhash has been extensively used
to detect potential duplicate content across websites in Google's
webpage crawler since 2006~\cite{GoogleCrawler}. Moreover, Google's
\floc proposal intended to use \simhash as a LSH for ad-targeting
purposes~\cite{FLOCWhitePaper}.

In \sysname, and similarly to \floc, we use \simhash to classify the
user's input to a specific category. Specifically, the classification
will be performed on the client's local device. The client profile will
be the input and the output of the \simhash will be mapped to a category
using pre-defined mapping.


\section{Technical Details}
\label{sec:construction}

There are three main entities in our system: 
\begin{itemize}
  \item the \emph{client} browsing websites:
  \item the \emph{advertiser} who wants to display ads on the client device;
  \item the proxy server, which acts as an intermediary between clients
  and the advertisers. 
\end{itemize}
The proxy hosts ads from different advertisers and the client fetches
these ads directly from the proxy. The proxy also simplifies preserving
the user privacy because clients never interact with the advertisers
directly. Throughout this section we will assume that the proxy and
advertisers are honest-but-curious, i.e. they follow the protocol and try to
learn extra information from the client messages and other side
channels. See Section~\ref{sec:threat} for more details about our threat
model.

\point{Private Data}
The main goal of our system is to hide a user's private data, while still
allowing the advertisers to show ads relevant to them. Unlike many previous
approaches, our system does not reveal any information
about the private data. In our system, any data related to a user's
browsing activity is considered private. This includes: visited
websites, content clicked, searched keywords, store browser cookies. Any
data that is extracted using the user's browsing activity is also
private. We say that protocols that leak none of this data to the proxy
provide \emph{full privacy}.

\begin{figure*}
  \centering
  \includegraphics[width=0.8\textwidth]{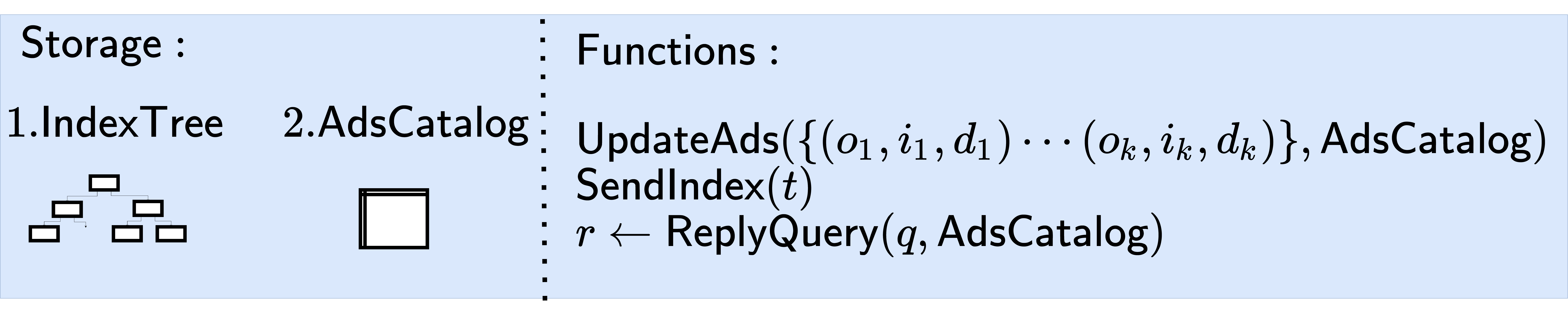}
  \caption{\sysname server's storage and function calls.}
  \label{fig:server}
\end{figure*}

\point{Server Proxy}
Figure~\ref{fig:server} shows the storage and API for the proxy server.
The proxy is assumed to enjoy large storage and computational capabilities.
In our system, we assume that the proxy is run by an independent entity
relative to the advertisers.
However, it is important to note that the security of our system assumes
an untrusted proxy that is attempting to learn more about clients based
on their queries.

The proxy stores two kinds of data in its storage.
\begin{itemize}
  \item {$\adcatalog$:} Dynamic storage for advertisements. The ads are indexed by
  categories. Each category consists of multiple ads. The catalog has
  can be thought of as a hierarchical structure with categories and
  sub-categories. Specifically, each ad is associated with a
  particular category and each category could, in turn, be linked to a
  category high up in the hierarchy.
  \item {$\indextree$:} At a high-level this data-structure maps
  categories to their indexes in the $\adcatalog$. The data
  structure is parsed as a tree. Each internal node in the tree consists
  of the category name and address of its children. The leaf nodes
  additionally contain category index in $\adcatalog$. This data
  structure provides a search operation that takes \emph{bit-string} as
  input and outputs node that could be reached by using the bit-string.
\end{itemize}

\begin{figure}[tb]
  \centering
  \includegraphics[width=1\columnwidth]{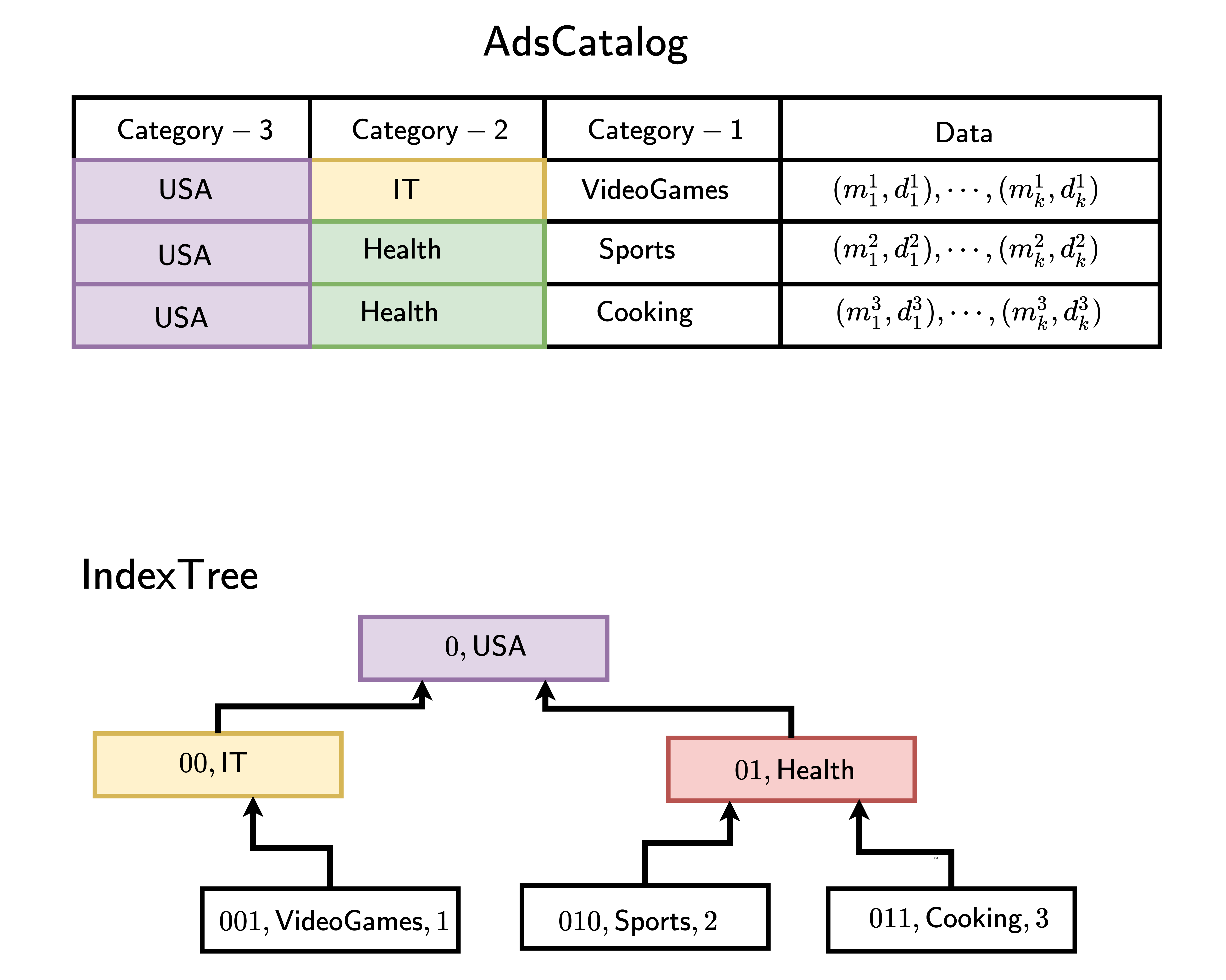}
  \caption{$\adcatalog$ and $\indextree$ data
  structures stored at the Proxy.}
  \label{fig:catalog}
\end{figure}

\point{Ad catalog} 
In Figure~\ref{fig:catalog} we give an example of such a data structure.
The example catalog consists of three categories each having $k$ relevant ads.
These ads could potentially belong to different advertisers. The catalog
has three levels of a category hierarchy. Note that each category level
is more \emph{targeted} than the higher level. For example, one of the
categories in level-1 is targeting people with an interest in cooking.
However, in level-2 the categories are targeting large population
interest in health or information technology. Similarly, at level-3
there is only one category targeting the whole US population. Note that
the exact relationship between the categories is not fixed and could be
decided by the proxy and the advertisers combined.

Figure~\ref{fig:catalog} also shows an example $\indextree$
based on the catalog. All the nodes in the tree could be accessed using
a bit string of size three.

\begin{figure*}
  \centering
  \includegraphics[width=0.6\textwidth]{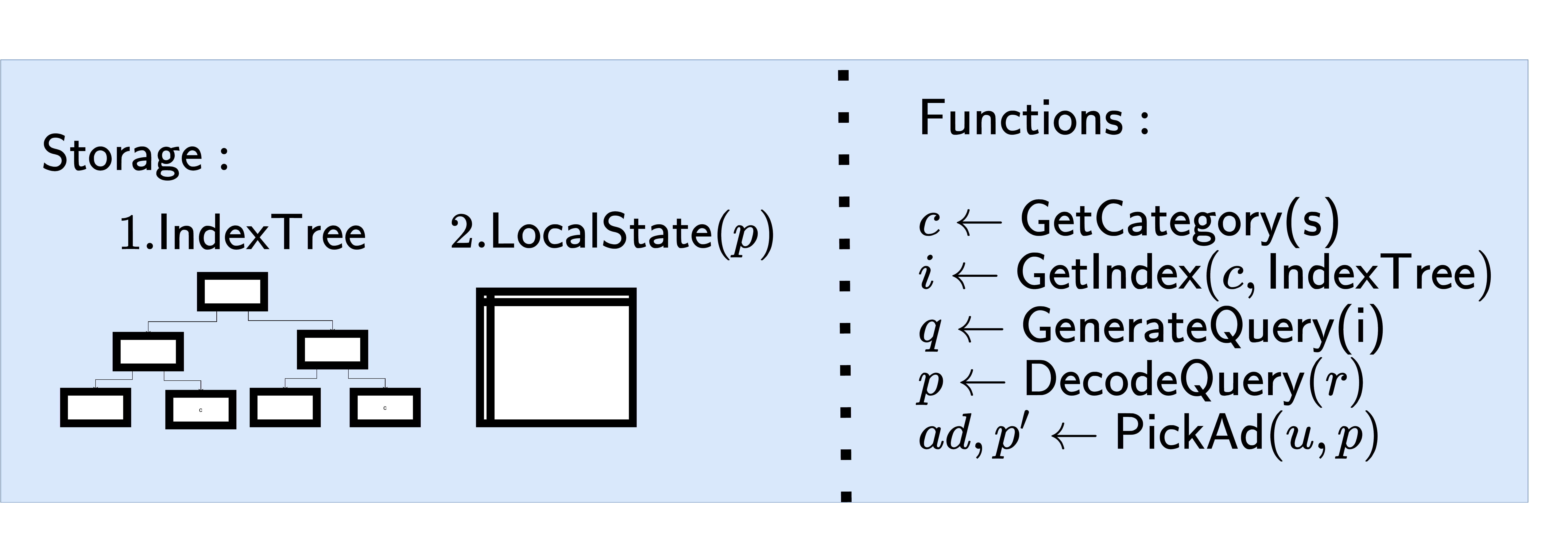}
  \caption{\sysname client's storage and function calls.}
  \label{fig:client}
\end{figure*}

\point{Client-proxy interaction}
Figures~\ref{fig:server} and \ref{fig:client} represent the storage and
the function calls used by the proxy and the client, respectively. The client stores
$\indextree$ and a small state $p$. The proxy on the other hand
stores the entire $\adcatalog$. Note that the size of
$\indextree$ is significantly smaller than
$\adcatalog$. For $k$ leaves, the total size of
$\indextree$ is $(2k-1) \log (2k-1)$ bits. Concretely, in our
implementation the size of $\indextree$ is approximately $3$ KB
while $\adcatalog$ consists of about $900$MB of data. We assume that the proxy
maintains an updated copy of $\indextree$, which any client
can download opportunistically.

\subsection{Protocol Flow}
\label{sec:flow}
\begin{figure*}
  \centering
  \includegraphics[width=1\textwidth]{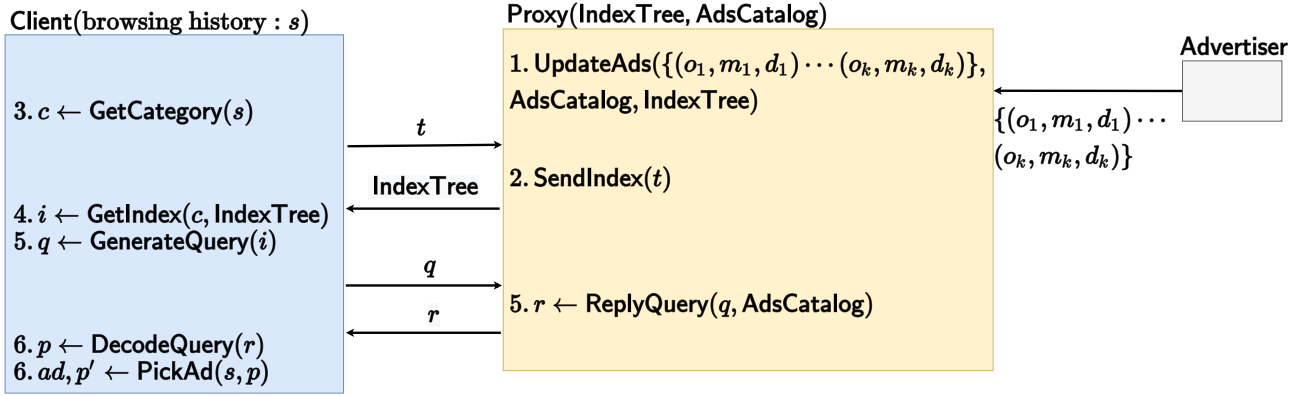}
  \caption{Overall flow of \sysname. The details are discussed in
  Section~\ref{sec:flow}}
  \label{fig:protocol}
\end{figure*}
In Figure~\ref{fig:protocol} we show the high-level flow of our system.
Here we define that flow in detail.
\begin{enumerate}
\item The advertisers upload their ads
$\{(o_1,m_1,d_1)\cdots(o_k,m_k,d_k)\}$ to the Proxy. Where each entry is
a tuple $(o_i,m_i,d_i)$, where $o_i$ is the operation {\texttt{add}, \texttt{remove},
\texttt{update}}, $m_i$ consists of ad identifier and the auction/matching logic,
and $d_i$ is the ad data. The advertisers could also specify the
category associated with each ad. In case of $o_i=\texttt{remove}$, $d_i$
will be empty and $m_i$ will contain ad id only. The proxy then calls
$\mathsf{UpdateAds}$ function to update the data structures.

\item The client will ask the proxy for the $\indextree$. To
achieve this the client will send the proxy \emph{last download time}
$t$ and the proxy will send back the most recent copy of the tree. If
the client has never downloaded the tree before, then it will send $0$
as the download time and the proxy will send the whole tree to the
client.  


\item The client will locally call a function $\mathsf{GetCategory}$ on her
preferences (e.g., browsing history) $s$ to get a relevant category. Like Google \floc this
step is implemented using the $\simhash$ function that takes as input
the preferences and outputs a category as a binary string.

\item The client will locally call $\mathsf{GetIndex}$ function with category
$c$ and $\indextree$ as inputs and get $i$ as output, which represents
the index of category in proxy's $\adcatalog$. If the client's category
is not present in $\indextree$, the function outputs the index of the
immediate sibling. Considering the example given in
Figure~\ref{fig:catalog}, on input category string of $000$ the
functions will output the index $1$ corresponding to the string $001$.

\item Using the index $i$ as input, the client will locally call function
$\mathsf{GenerateQuery}$ and sends the output query $q$ to the proxy.
The proxy will then call function $\mathsf{ReplyQuery}$ with inputs $q$
and $\adcatalog$ and send the output $r$ to the client.
$\mathsf{GenerateQuery}$ function directly calls $\pirscheme.\query(i)$
and $\mathsf{ReplyQuery}$ function directly calls
$\pirscheme.\reply(q)$.

\item The client then locally calls $\mathsf{DecodeQuery}$ function,
based on $\pirscheme.\extract$, to retrieve a set of ads from the proxy
response. The client then stores these ads as a local state $p$.

\item Finally the client calls $\mathsf{PickAd}$ function, which picks
the relevant ad from the state. This function takes into
consideration the matching logic associated with each ad. Note that
each call to this function consumes some part of state $p$ as ads are
displayed to the user. 
\end{enumerate}


\section{Evaluation}
\label{sec:evaluations}
In this section, we report on the performance and scalability of
\sysname in terms of bandwidth, \mbox{client-side} computation and
\mbox{server-side} computation. We measure the effectiveness of our
system to meet the performance requirements laid out in
Section~\ref{sec:intro}. Such metrics include the size of the server's
database; the number of concurrent queries from the client; the required
bandwidth usage; and, finally, associated the monetary cost of
implementing such a system in common hardware. In the ad catalog, each
category consists of $30$ relevant ads. Also, each client query results
in fetching all of $30$ ads. The client then consumes these ads from the
local state. Note that this design choice considerably improves the
overall performance of \sysname because the cost of each client-server
interaction is amortized over $30$ ads. We assume that per day each
client sends $10$ queries. It means that each client is shown around
$300$ ads per day. We ran our experiments~$10$ times and report their
averages below.

\subsection{Bandwidth Evaluation}
\begin{figure}[t]
  \setlength{\tabcolsep}{12pt}\small\centering
  \begin{tabular}{rr}
  \toprule
  \bf Number of Ads & \bf Communication Size (KB) \\
  \midrule
  \(262,144\) & \(192\) \\
  \(1,048,576\) & \(192\) \\
  \(4,194,304\) & \(192\) \\
  \(16,777,216\) & \(192\) \\
  \bottomrule
  \end{tabular}
  \caption{Size of ad fetching query and response as a function of the number of items in the database.}
  \label{query-size}
\end{figure}
Figure~\ref{query-size} represents the total communication volume
between the client and the server. In \sysname the communication size is
independent of the database size. In total, the communication size
is~$192$~KB, which includes~$64$~KB of data from the client to the
server and~$128$~KB of data from the server to the client. This
communication is mainly due to the query and response size of OnionPIR.
OnionPIR has the smallest communication volume among all the
single-server PIR schemes. However, to further improve the communication
bandwidth, we could utilize multi-server PIR schemes. We avoid adopting
that architecture as it requires distributed trust among the multiple
servers, which is not a suitable setting for ads. 

We further believe that the small communication size of our system makes
it an ideal candidate for bandwidth-constrained clients, such as mobile
phone users.

\subsection{Runtime Latency}
\begin{figure}[t]
  \centering
  \setlength{\tabcolsep}{12pt}\small\centering
  \begin{tabular}{rr}
  \toprule
  \bf Number of Ads & \bf Computation (sec) \\
  \midrule\small \(262,144\) & \(10\) \\
  \(1,048,576\) & \(40\) \\
  \(4,194,304\) & \(160\) \\
  \(16,777,216\) & \(640\) \\
  \bottomrule
  \end{tabular}
  \caption{Server-side computation time to generate a response. For a database with one million entries, the server only takes only~$40$ seconds to generate a reply.}
  \label{compute-server}
\end{figure}
In Figure~\ref{compute-server}, we present the time it takes for the
server to deliver ads relevant to a single client. Our system is capable
of delivering \(30\) ads in a given category in around~$10$ seconds,
even when $\adcatalog$ has more than~$250,000$ categories. In our
experiments, we also found that around~$90\%$ of computation is due to
underlying PIR operations. Without using a different PIR scheme,
improving this further requires improving underlying crypto primitives.
While improvement may potentially be made using alternative PIR schemes
in multi-server or stateful models, we highlight issues with these
approaches in Section~\ref{sec:discussion}.

\subsection{Server Cost for Running \sysname}
\begin{figure}[t]
  \setlength{\tabcolsep}{12pt}\small\centering
  \begin{tabular}{rr}
  \toprule
  \bf Number of Ads & \bf Server Cost (US cents) \\
  \midrule
  \(262,144\) & \(0.83\)\\
  \(1,048,576\) & \(3.33\)\\
  \(4,194,304\) & \(13.33\)\\
  \(16,777,216\) & \(53.33\)\\
  \bottomrule
  \end{tabular}
  \caption{Monthly cost of each user in \sysname. We assume that the user make~$10$ queries for buckets of~\(30\) ads each day.}
  \label{client-cost}
\end{figure}
In Figure~\ref{client-cost}, we calculated monthly per-user monetary
cost of running \sysname. The computation and network cost of an Amazon EC2
\texttt{t2.2xlarge} instance, rented at \(\$0.01\) per core hour\footnote{\url{https://aws.amazon.com/ec2/spot/pricing/}} and nine
cents per gigabyte of data transfer.\footnote{\url{https://aws.amazon.com/ec2/pricing/on-demand/}} 

Consequently, in \sysname per user monthly bandwidth is only~$0.005$
cents. In terms of computation, the monthly per-user cost of serving an
ad bucket from a $\adcatalog$ of size~$262,144$ for a user that makes 10
queries of 30 ads per day is \(0.83\) cents. According to some published
estimates~\cite{gads}, Google's monthly ad revenue is
approximately~5~USD per user. This means that Google, using \sysname,
would operate with a $1\%$ profit margin relative to their per-user
revenue, whilst serving ads in a completely privacy-preserving manner.

\point{Financial comparisons}
We compare the financial costs of running \sysname with the trivial
solution of sending the entire ad database to each client. We assume
bandwidth usage costs based on retrieving a new ads database every two
hours, where around \(65\%\) clients are online at any one time. This
mirrors the deployment scenario used in~\cite{BraveAdCatalog}. Overall,
we expect the total cost to amount to~\(8.75\) cents per client. This
cost arises from assuming that each client downloads the ad an ad
database of approximately \(0.25\)GB (\(1\)KB ads \(\times\)
\(262,144\) entries). Thus, the financial costs of running \sysname
are a \(10\times\) reduction compared with the trivial solution.
Moreover, this improvement scales identically as the database increases.

Finally, we note that the eventual cost of running \sysname is at least
\(4\times\) larger than the concurrent work of
AdVeil~\cite{EPRINT:SerHogDev21}.\footnote{Costs are extrapolated
somewhat due to differing hardware usage and thus this is only an
approximation.} However, AdVeil also provides categorically weaker
privacy guarantees, and relies on expensive anonymizing proxies such as
Tor. We provide a more detailed comparison with all related work in
Section~\ref{sec:related}.

\section{Discussion}
\label{sec:discussion}
\point{Reporting ad conversions privately}
Ad interaction reporting is an important aspect of the online
advertisement ecosystem. An accurate ad interaction reporting is
essential to ensure that i) advertisers are billed as a function of the
number of interactions their ads have over time; and that ii) publishers
are paid fairly. With \sysname, the ad distribution party does not have
visibility of how users are interacting with ads in the system. In
addition, the users should not trivially report the ad interaction to
the advertisers and ad distribution party. Doing so would render useless
the efforts to protect the user privacy by hiding the ad fetching
request patterns with \sysname. Thus, when using our system, it is
impossible for the ad distribution party to assemble an ad interaction
report that can be used to bill advertisers.

However, there are multiple protocols that can be deployed in parallel
with \sysname to provide privacy preserving ad reporting.
Adnostic~\cite{NDSS:TNBNB10} proposes a system based on homomorphic encryption and
\mbox{zero-knowledge} proofs to provide secure and private ad reporting.
The authors of Privad~\cite{NSDI:GuhCheFra11} introduce an entity called the Dealer that
is responsible for anonymizing user interactions with ads and respective
billing. In~\cite{privreporting}, the authors leverage an additively encryption
scheme that enables privacy-preserving ad reporting at scale. Finally,
THEMIS~\cite{ARXIV:PQPL21} proposes a private ad reporting mechanism based on an
homomorphic encryption, a threshold signature scheme, and a
\mbox{peer-to-peer} network. In terms of existing practical deployments
of privacy-preserving reporting ad conversions, the Safari browser 
currently tracks conversions locally and then report these to a server
without revealing the user identity~\cite{AppleConversions}.

All of the approaches highlighted above could be used as a
privacy-preserving reporting layer that embeds the \sysname as the
targeting and delivery layer, without impacting any of our original
privacy guarantees.

\point{Real-time advertisement auctions}
Unlike systems such as AdVeil~\cite{EPRINT:SerHogDev21}, we are unable
to build real-time advertisement auctions into the ad-delivery pipeline.
This is only possible in previous work because the untrusted proxy that
is used is able to see which ads are queried, and unlinkability of ad
views to user profiles is maintained via an \emph{anonymizing proxy}
(see Section~\ref{sec:related} for more details).

In \sysname, the untrusted proxy learns nothing from a client query.
Therefore, ad auctions are only possible to the extent that ad groupings
for the locality-sensitive hash function that is used (\simhash) can be
decided apriori. While this is a regression when compared with other
systems, we believe that the increased user privacy in our system is of
paramount importance. Moreover, in theory such auctions may be possible
to run within FHE circuits, although such capabilities are still far
beyond the realm of practical web applications.

\point{Alternative PIR schemes}
In \sysname, we do not use these approaches due
to their inherent assumptions that are not suitable for the application.

\begin{itemize}
    \item \textsc{Stateful PIR}: Patel et al.~\cite{CCS:PatPerYeo18} introduced single-server stateful
    PIR where the client retrieves some helper data in the offline phase and
    use it to make the online PIR queries. Their protocol has substantially
    reduced the amortized computation cost over vanilla stateless PIR.
    However, their scheme requires the client to download the entire
    database in the offline phase. For applications like online advertising,
    where the database is potentially large, it is impractical to download
    the entire database.

    \item \textsc{Batched PIR}: Batched PIR allows the server to answer a batch of PIR queries
    at a lower cost than answering each query separately. This general
    strategy is adopted by various protocols~\cite{IshaiKOS04, Henry16,
    SP:ACLS18, JC:BeiIshMal04, IshaiKOS06}. We remark that this approach is
    not always applicable in \sysname because the client access only one
    index at a given time.
    
    \item \textsc{PIR with Pre-processing}: Another direction is PIR with pre-processing, first proposed by Beimel et
    al.~\cite{JC:BeiIshMal04}. In their scheme, the server first performs a
    linear pre-processing step; after that, the server's work per query is
    sub-linear. However, this scheme has an exponentially large storage
    overhead for the server. In other words, their scheme requires the
    server to store a separate copy of the pre-processed database for each
    client.
\end{itemize}

\point{Preventing malicious server activity}
While we only consider a honest-but-curious server in our construction, we note
that a malicious server could arbitrarily alter their input to try and
learn more about the client queries. An example of such an attack would
see the server host \emph{bad} files at certain points in the database,
that cause the client to abort the protocol in some way that the server
can recognize. The server may then be able to try and leak which
indices the client is querying by carefully choosing which files to
corrupt, and then inspecting whether an abort occurs.

One way around this is for the server to prove to the client in
zero-knowledge that each entry of the database is well-formed, although
such a solution is likely to be prohibitively expensive for the client.
A more practical alternative is for the server to either have
its database input certified and signed by a third-party, and then
verifying that the server input is signed correctly before
proceeding.\footnote{In addition, one could avoid verifying a signature
over each record using a cut-and-choose approach that only verifies a
random selection of records.} Such a solution could also be integrated
with a public verification procedure, where clients are able to inspect
the public database at an independent trusted location, and then
verifying that the server input is the same as the one that it has seen
previously.\footnote{Note that, in our application (and PIR in general),
the server database is considered to be public information, and so this
does not impact any security guarantees.} Finally,
implementation-specific countermeasures could include careful handling
of client aborts, to make sure that they are indistinguishable to the
server from client successes.

\point{Other content-fetching applications}
Similarly to online advertising systems, there is a  wide range of web
applications that require users to fetch content from third parties.
From revoked certification checks performed by web
browsers~\cite{SP:LCLMMW17}, to map\footnote{https://maps.google.com}
and weather applications\footnote{https://www.accuweather.com}, the
browser is often required to issue queries to \emph{third-party}
providers that leak information about the user's profile (such as cookie
state) and behavioral aspects. For example, when the browser fetches the
forecast from a weather service, the service provider learns the
location that the user is interested in; this is likely the location
where the user is or plans to be.

Although we focused on the particular case of online advertising in this
work, our content fetching system is positioned to be a practical and efficient
solution to provide \emph{privacy-preserving} content fetching capabilities to 
any web application. \sysname can be used as a \mbox{drop-in} replacement to any
content fetching protocol implemented by web applications to add strong privacy guarantees
to users. In order to integrate \sysname in an existing web application, the service provider
needs to run a \sysname proxy as described in Section~\ref{sec:construction}. In addition,
the client must integrate the \mbox{high-level} APIs to generate and handle the \pir queries,
namely $(\pparams) \gets \pirscheme.\init(c, \db):$, $(q) \gets \pirscheme.\query(j,\state)$,
and $(\mathsf{B}) \gets \pirscheme.\extract(r, \state)$ described in
Section~\ref{subsec:pir_framework}. These routines can be implemented in JavaScript and
easily integrated across different web applications.

\section{Related Work}
\label{sec:related}
\subsection{Private Information Retrieval}
\label{sec:pir-related}
As mentioned previously, private information retrieval protocols are
proposed in single-server and multi-server models.

The first single-server scheme was proposed by Kushilevitz and
Ostrovsky~\cite{FOCS:KusOst97}. In their scheme, the database is
represented as a high-dimensional hypercube. the client's request is
encrypted under additive homomorphic encryption. The scheme has a
request size of $O(\sqrt{N} \log N)$ and response size of $O(\sqrt{N})$.
Later on, several works have extended this scheme using different
cryptographic assumptions. For example, Cachin et al.~\cite{CachinMS99}
proposed a PIR protocol based on $\phi$-Hiding assumption,
Chang~\cite{Chang04}' scheme is based on Pailer homomorphic encryption
and Lipmaa~\cite{Lipmaa05} uses the Damgard-Jurik encryption
scheme~\cite{DamgardJ01}. All of these schemes have improved the request
and response size of the original protocol. However, it has been
observed that these schemes in practice often perform slower than
downloading the entire database when the network bandwidth is a few
hundred kilobytes per second~\cite{sion2007computational}.

Recently more practical single-server schemes have been proposed that
display promising performance for various applications. Aguilar-Melchor
et al.~\cite{PoPETS:ABFK16} present XPIR with good computation cost.
But, their protocol has a very high request and response size.
SealPIR~\cite{SP:ACLS18} addresses the request size bottleneck by
introducing a novel query compression technique. This results in a
significant reduction in request size however their response size is
similar to XPIR. Ali et al.~\cite{AliLP0SSY19} gives a protocol that
improves upon SealPIR's response size. However, their scheme has a
higher computational overhead than SealPIR and XPIR.  Recently, Mughees
et. al.~\cite{EPRINT:MugCheRen21} has proposed OnionPIR. OnionPIR has
significantly reduced the response overhead of SealPIR while keeping the
computation comparable. Concretely, the response overhead is only $4.2$x
over the insecure baseline. 

Due to its performance advantage over previous schemes, we have used
OnionPIR as the underlying PIR scheme for \sysname. The OnionPIR scheme
excels in terms of the request and response sizes (i.e., communication)
but the computation overheads remains quite high. In all of these
schemes, the server needs to perform at least $N$ expensive
cryptographic operations and the computational cost of such an operation
is often higher. As mentioned in Section~\ref{sec:pir-background}, we
could use alternative PIR schemes to reduce computation overheads, but
doing so introduces significant costs in terms of communication,
implementation, and management of state and database updates (Section~\ref{sec:discussion}).

\subsection{Private Advertising}
\begin{figure*}[t]
    \setlength{\tabcolsep}{12pt}\small\centering
    \scalebox{0.9}{%
        \begin{tabular}{lllll}
        \toprule
        \bf Protocol & \bf Accuracy & \bf Leakage & \bf Trusted Third Party & \bf Financial cost \\\midrule

        Privad~\cite{NSDI:GuhCheFra11} & Broad interest categories & Ads without client identifiers & Yes & Negligible \\
        \floc~\cite{FLOC} & Broad interest categories & Broad interest categories & Yes & Negligible \\
        AdVeil~\cite{EPRINT:SerHogDev21} & Fully targeted & Ads without client identifiers & Yes (Anonymity Proxy) & $<0.75$ cents \\
        ObliviAd~\cite{SP:BKMP12} & Fully targeted & None & Yes (TEE) & TEE costs \\
        \hline
        Adnostic~\cite{NDSS:TNBNB10} & Contextually Targeted & Contextual & No & Negligible \\
        Brave Ads~\cite{BraveAdCatalog} & Fully targeted & None & No & $\sim$30 cents \\
    
        \sysname & \textbf{Fully targeted} & \textbf{None} & \textbf{No} & \textbf{$\sim$3 cents} \\
        \bottomrule
        \end{tabular}
    }
    \caption{
        Comparison with related work in privacy-preserving
        advertisement targeting and delivery. All cost estimates relate to
        serving clients from a database of 1 million 1KB advertisements from
        the EC2 hardware that we note in Section~\ref{sec:evaluations}. Note
        that "Fully Targeted" refers to both behavioral and contextual
        targeting.
    }
    \label{table:comparison}
\end{figure*}
There have been multiple attempts to design and implement advertising
ecosystems that are practical and privacy-preserving. In the table in
Figure~\ref{table:comparison}, we compare the characteristics of
\sysname with approaches taken from both previous research in this area
and real-world deployments. We discuss these systems in further detail
below.

In~\cite{RSA:Juels01}, the authors introduce the concept of private
targeting advertising on the web and propose \mbox{\pir-based} systems
to deliver ads privately. This work is mostly of historical interest,
since the theoretical \pir schemes described there are neither
practical, nor do they impose favorable trade-offs in terms of user
privacy; they also rely on an external mixnet networks to achieve user
privacy.

Adnostic~\cite{NDSS:TNBNB10}, proposes a system that features behavioral
ad targeting and user privacy and is complementary to the current web ad
infrastructure. The user fetches a set of ads from the ad Broker and the
behavioral targeting happens locally. The browser processes the user's
history to determine their interests, which are then used to select a
subset of the fetched ads to show the user. In addition, Adnostic
proposes a cryptographic scheme that relies on homomorphic encryption
and \emph{zero-knowledge} proofs to implement privacy-preserving billing
reporting system. This system allows publishers and advertisers to learn
the performance of the ad campaigns while preserving user's privacy.
However, Adnostic does not attempt to hide which ads the user requests
from the \emph{Broker}, which leaks user interests and behavior to third
parties.

Privad~\cite{NSDI:GuhCheFra11} addresses the privacy concerns in the web
advertising industry by introducing a new party~--~the
\emph{Dealer}~--~that proxies and anonymizes all interactions between
the user and the ad providers. The communication between the user and
the Dealer is encrypted, thus the Dealer does not learn any behavioral
information about the users. On the other hand, the ad providers have
access to the user's behavioral information but do not know the user's
identity. In addition to providing privacy, the Dealer is also
responsible for the billing logic and fraud prevention. The drawbacks of
Privad are that the Dealer is a central party that needs to be online
and intermediate all the communication between the users and ad
providers. In addition, the user's privacy requires \emph{non-collusion}
between the Dealer and ad providers.

The authors of ObliviAd~\cite{SP:BKMP12} propose an
\emph{hardware-based} PIR system that provides strong security and
privacy through an Trusted Execution Environment~(TEE) and on an
Oblivious RAM (ORAM) scheme~\cite{STOC:Goldreich87}. The TEE ensures
that the ad targeting, ad billing reporting and fraud prevention are
\mbox{privacy-preserving}. The user sends an encrypted behavioral
profile to a third party running a TEE environment. The ad targeting
logic runs on the TEE and selects a batch of ads based on the encrypted
user profile. Finally, ObliviAd relies on an ORAM scheme to ensure that
the ad fetching does not leak sensitive information about the user.
Although ObliviAd provides strong security and privacy guarantees, it
assumes that the TEE environment is secure against attacks that may
expose the privacy of user’s data being processed within the TEE. As it
has been shown in recent literature, sadly, there are no instances of
TEE designs that provide those guarantees~\cite{SP:CSFP20}.

THEMIS~\cite{ARXIV:PQPL21} is a decentralized and
\emph{privacy-preserving} ad platform that provides auditability,
rewards users for viewing ads, and allows advertisers to verify the
performance and billing reports of ad campaigns. The user privacy at ad
targeting and fetching phases is guaranteed by relying on a trivial PIR
scheme and by performing local ad targeting. First, the user downloads
the whole database periodically from a third party who curates the ad
catalog. Then, the user selects locally the ads to view based on their
profile and browsing history. Since no user-specific data leaves the
client, the system does not leak any sensitive information about the
user. In addition, THEMIS relies on homomorphic encryption and
\emph{zero-knowledge} proofs to protect the user privacy at ad
accounting and billing phases. Due to relying on a trivial \pir scheme,
a major limitation of THEMIS is to scale as the number of ads in the
catalog increase given the bandwidth necessary for all users to download
the whole ad database. \sysname could be used as a \emph{drop-in}
replacement for the trivial \pir scheme to overcome this limitation.

In concurrent work, AdVeil~\cite{EPRINT:SerHogDev21} proposes a modular
\emph{privacy-preserving} advertising ecosystem with formal guarantees
for end users. The system features private ad targeting, private ad
retrieval, and a private ad reporting scheme. The private ad targeting
subsystem relies on a single-server \pir protocol and a
\mbox{locality-sensitive} hashing mechanism to allow the users to learn
which ads to fetch from the broker without disclosing their profile.
Both the private ad retrieval and the reporting scheme rely on an
anonymizing proxy (e.g. Tor) to ensure the unlinkability between the
user's preferences and the queries issued to the ad broker. Although the
fetching performance achieved by AdVeil is better than \sysname, the
biggest drawback is its reliance on an anonymity proxy to protect the
users' privacy when fetching ads from the broker. Using anonymous
proxies correctly such as Tor is not trivial for average web
users~\cite{FGKMN10,ACM:ClaOorAda07,NBCC14}. In addition, many ISPs and
private networks block access to such networks, which effectively
prevents AdVeil users from successfully fetching and displaying
ads~\cite{IES:SapNadBar16, KFAJSPMM16, MilCurLun15}. \sysname, on the
other hand, relies on \pir for \emph{both} the ad targeting and ad
fetching phases, which provides stronger privacy and usability
guarantees without relying on external systems such as anonymity
proxies.

\section{Conclusion}
\label{sec:conclusion}
In this work, we devise a new framework called \sysname for content
delivery that relies on cryptographic hardness and best-case privacy,
rather than syntactic and optimistic privacy guarantees. In \sysname,
our scheme utilizes local computation of preferences followed by
efficient, configurable, single-server private information retrieval
(PIR) to ensure that clients can fetch content from servers, without
revealing any of their inherent characteristics to the content provider.

Our solution works by combining novel cryptographic
optimizations to PIR schemes that allow storing minimal client state, in
order to gain better practicality, that should have independent benefit
to the building of practical PIR schemes.
In the context of advertisement delivery, we show that we can
deliver~\AdCount ads to a client in~\Runtime seconds, with total
bandwidth costs of~\Comms{}MB, where the ad database has~\DBSize
entries. In a system with~\ClientCount clients, we calculate that using
\sysname gives a \(10\times\) financial saving in delivering such content to
users, compared with the trivial solution of sending the entire
database. In addition, performance is comparable with similar ad
targeting mechanisms that provide weaker guarantees with regards to
leakage of client query information.

Overall, our results show that practical advertisement targeting and
delivery systems with best-case privacy guarantees can be built from PIR
protocols. We expect that \sysname can be used to deliver the new
state-of-the-art in privacy-preserving behavioral ad targeting and
delivery.  Moreover, we believe that our methods could be applied to
other settings.

\point{Future Work} In \sysname we have utilized single-server
\emph{stateless} PIR protocol. One way to directly improve the latency
of the system is to harness the power of stateful or
batched frameworks for running single-server PIR~\cite{IshaiKOS04, Henry16,
SP:ACLS18, JC:BeiIshMal04, IshaiKOS06, CCS:PatPerYeo18}. However, as discussed in
Section~\ref{sec:discussion}, these schemes are not directly compatible
with \sysname, therefore an interesting future direction is to extend
\sysname to make it possible to embed these approaches as the underlying
PIR scheme.  

\Urlmuskip=0mu plus 1mu\relax
\bibliographystyle{IEEEtran}
\bibliography{./bib/local,./bib/abbrev3,./bib/crypto}
\end{document}